\documentclass[11pt]{article}

\usepackage{amsmath}
\usepackage{graphicx}
\usepackage{amsfonts}
\usepackage{amssymb}
\usepackage{epsfig}
\usepackage{color}
\usepackage{psfrag}
\usepackage{epstopdf}

\newcommand{\EQ}{\begin{equation}}
\newcommand{\EN}{\end{equation}}
\newcommand{\be}{\begin{equation}}
\newcommand{\ee}{\end{equation}}
\newcommand{\bea}{\begin{eqnarray}}
\newcommand{\eea}{\end{eqnarray}}

\begin{document} \setcounter{page}{0}
\topmargin 0pt
\oddsidemargin 5mm
\renewcommand{\thefootnote}{\arabic{footnote}}
\newpage
\setcounter{page}{0}
\topmargin 0pt
\oddsidemargin 5mm
\renewcommand{\thefootnote}{\arabic{footnote}}
\newpage
\begin{titlepage}
\begin{flushright}
\end{flushright}
\vspace{0.5cm}
\begin{center}
{\large {\bf Interface localization near criticality}}\\
\vspace{1.8cm}
{\large Gesualdo Delfino}\\
\vspace{0.5cm}
{\em SISSA -- Via Bonomea 265, 34136 Trieste, Italy}\\
{\em INFN sezione di Trieste}\\
{\em E-mail: delfino@sissa.it}\\
\end{center}
\vspace{1.2cm}

\renewcommand{\thefootnote}{\arabic{footnote}}
\setcounter{footnote}{0}

\begin{abstract}
\noindent
The theory of interface localization in near-critical planar systems at phase coexistence is formulated from first principles. We show that mutual delocalization of two interfaces, amounting to interfacial wetting, occurs when the bulk correlation length critical exponent $\nu$ is larger than or equal to 1. Interaction with a boundary or defect line involves an additional scale and a dependence of the localization strength on the distance from criticality. The implications are particularly rich in the boundary case, where delocalization proceeds through different renormalization patterns sharing the feature that the boundary field becomes irrelevant in the delocalized regime. The boundary delocalization (wetting) transition is shown to be continuous, with surface specific heat and layer thickness exponents which can take values that we determine.
\end{abstract}
\vspace{.5cm}
\end{titlepage}

\newpage
\section{Introduction}
When a statistical system is at a point of phase coexistence, an interface separating two different phases can be induced imposing suitable boundary conditions. If, in addition, the system is close to a second order phase transition point (e.g. the Ising model slightly below the critical temperature) the interface exhibits universal properties whose characterization represents a particularly interesting theoretical problem. While in principle the field theory which describes the universal properties of the order parameter also encodes those of the interface, in practice the fact that the latter is an extended object makes non-trivial its study within local field theory (see e.g. \cite{ID}). 

A very relevant aspect is that of the localization of an interface at a boundary, on a defect, or with respect to a second interface. While the problem is clearly of general theoretical interest, a specific terminology has been developed in the context of wetting phenomena, the name referring to a liquid-vapor interface, with the liquid phase wetting the boundary upon delocalization of the interface (see e.g. \cite{Dietrich,BEIMR}). In this case, however, the long range forces normally relevant in fluids move the focus away from the universal properties we are interested in. In this paper we will only consider systems with short range interactions close to (second order) criticality, for which the problem of interface localization can be addressed within the framework of universality, and then of statistical field theory. The wetting vocabulary, on the other hand, is traditionally used also for short range interactions, and we will often refer to it in this paper.

We will consider the two-dimensional case and show how interface localization can be analyzed within the field theory describing the universality class of the order parameter. As will become clear, the main reason why we are able to achieve this goal is that in two dimensions the interfaces correspond to trajectories in imaginary time of the topological excitations (kinks) of the underlying (1+1)-dimensional quantum field theory. Extracting the implications of this basic circumstance we arrive, in particular, at the following results. Mutual delocalization of two interfaces, also known as interfacial wetting, occurs when the correlation length critical exponent $\nu$ is larger than or equal to 1. Localization on a defect line persists arbitrarily close to criticality, despite the fact that its strength tends to zero if the defect operator has scaling dimension larger than 1. An interface delocalizes from a boundary sufficiently close to criticality, and in the delocalized regime the boundary field becomes irrelevant in the renormalization group sense. This delocalization transition is continuous (within the classification usually adopted in this context), with allowed values of the interfacial specific heat exponent $\alpha_S=-4k$, $k=0,1,2,\ldots$; for the layer thickness exponent we obtain $\beta_S=\alpha_S/2-1$. The simplest case $k=0$ explains the exponents known from the lattice solution of the Ising model \cite{Abraham,Abraham_wetting}. 

The paper is organized as follows. The next section is devoted to the field theoretical description of the bulk theory and to its implications for the mutual localization of two interfaces. Sections 3 and 4 are then devoted to localization by a defect line and a boundary, respectively, while the last section contains a short summary.

\section{Mutual localization of two interfaces}
We will consider two-dimensional statistical systems with values of the bulk parameters corresponding to coexistence of different phases. In particular, we are interested in the regime in which such a system is close to a second order phase transition point, to which we refer in the following as the critical point. This means that the bulk correlation length $\xi$ is much larger than microscopic scales\footnote{At the same time $\xi$ is much smaller than the linear size $R$ of the system, which is then regarded as infinite in the following. The condition $R\gg\xi$ is necessary for the obervation of separated phases.} and that the system admits a continuous description in terms of a field theory specified by a reduced Hamiltonian (or Euclidean action)
\EQ
{\cal H}_\textrm{bulk}={\cal H}_\textrm{critical}+\lambda\int d^2x\,\Phi(x)\,,
\label{bulk}
\EN
where ${\cal H}_\textrm{critical}$ is the scale-invariant reduced Hamiltonian of the critical point, $\lambda$ measures the distance from criticality, and $\Phi(x)$ is the operator which drives the system away from criticality.
The fact that we are at phase coexistence in two dimensions ensures that $\xi$ is finite for $\lambda\neq 0$, and that the operator $\Phi(x)$ is relevant (or marginally relevant) in the renormalization group sense, with a scaling dimension $X_\Phi\leq 2$, the equality corresponding to the marginally relevant case. It follows from (\ref{bulk}) that 
\EQ
\lambda\sim\xi^{X_\Phi-2}\,,
\label{lambda}
\EN
or conversely $\xi\sim\lambda^{-\nu}$ with $\nu=1/(2-X_\Phi)$. 
The two-dimensional Euclidean field theory (\ref{bulk}) is the continuation to imaginary time of a relativistically invariant quantum field theory in one spatial dimension. Within the quantum description, the coexisting phases of the statistical system correspond to degenerate vacuum states $|0_a\rangle$, with $a=1,2,\ldots$ labeling the different phases. In a (1+1)-dimensional quantum field theory with degenerate vacua the elementary excitations have a topological nature, and correspond to the kinks $a|b$ connecting a vacuum $a$ to a vacuum $b\neq a$. Being a relativistic particle, a kink $a|b$ carries energy and momentum $(e,p)=(m_{ab}\cosh\theta,m_{ab}\sinh\theta)$, where $m_{ab}\sim 1/\xi$ is the kink mass and $\theta$ is the rapidity parameter. 
If, for two given phases $a$ and $b$, the elementary kink $a|b$ connecting the corresponding quantum vacua exists, its trajectory\footnote{The kinks are the excitations of the theory (\ref{bulk}) which describes all fluctuations near criticality. The notion of kink trajectory is intended within the field theoretical framework which sums over all possible configurations. It was shown in \cite{DV,DS2,fpu} how this determines the internal structure of the interface.} in imaginary time yields an interface separating phases $a$ and $b$ in the statistical system, with $m_{ab}$ exactly equal to the interfacial tension \cite{DV,DS2,fpu}. If $a|b$ does not exist, instead, going from $|0_a\rangle$ to $|0_b\rangle$ necessarily requires a multi-kink excitation, say a two-kink one $a|c|b$ yielding two interfaces enclosing a macroscopic layer of a third phase $c$ (Fig.~1a). Hence we see that this latter phenomenon, known as {\it interfacial} wetting, is actually determined by the vacuum connectivity structure of the underlying quantum theory. As we now explain, this in turn depends on the value of the critical exponent $\nu$. 

The problem can be restated as follows. Given a two-kink state $a|c|b$, there is interfacial wetting only if the two kinks do not allow for a stable bound state $a|b$ which, by definition of bound state, would have a mass $m_{ab}$ smaller than the total mass $m_{ac}+m_{cb}$ of the state $a|c|b$, and would lead, via free energy minimization, to a single interface along which phase $c$ is confined in a thin layer. The existence of a bound state is a property which does not change along the renormalization group trajectory defined by (\ref{bulk}), and can then be investigated in the tail of the trajectory, where particle kinetic energies are much smaller than their masses and non-relativistic potential theory applies. In particular, we can use the result that an attractive potential on a line produces at least a bound state \cite{Landau}. On the other hand, it was shown in \cite{DS2,DS5} through the exact study of the unbound regime that the kinks have fermionic statistics which accounts for the mutual avoidance of the interfaces. With reference to (\ref{bulk}), non-interacting fermions correspond to a Hamiltonian bilinear in the fermions, where $\Phi$ provides the mass term and $X_\Phi=\nu=1$ (a single fermion species corresponds to the Ising model, with $\lambda\sim T_c-T$). Hence the attractive regime corresponds either to $X_\Phi<1$ or to $X_\Phi>1$. A simple way to fix the issue is to consider the three-state Potts model. Indeed, given a two-kink state $a|c|b$, the permutational symmetry of the model implies the existence of the bound state $a|b$ with the same mass of $a|c$ and $c|b$. Since this model has $X_\Phi=4/5$ \cite{Nienhuis}, the binding regime corresponds to $X_\Phi<1$. Conversely, interfacial wetting occurs for $X_\Phi\geq 1$, i.e. $\nu\geq 1$. 

A basic illustration of this general result is provided by the Blume-Capel model \cite{Blume,Capel}, i.e. an Ising model in which non-magnetic sites (vacancies) are also allowed. As the temperature is lowered the model exhibits an ordering transition which is continuous up to a vacancy density $\rho_c$, and becomes first order above $\rho_c$. The first order line, along which the ferromagnetic phases $+$ and $-$ coexist with the disordered phase $0$, corresponds to (\ref{bulk}) with ${\cal H}_{\textrm{critical}}$ describing the tricritical point at $\rho_c$, $\lambda\sim\rho-\rho_c$, and $X_\Phi=6/5$ \cite{BPZ}. Since $\nu>1$, the result we obtained above implies that the state $+|0|-$ does not bind, and that a wetting layer of the disordered phase forms in-between the ferromagnetic phases. The absence of bound states can be checked within the exact scattering solution \cite{Sasha,dilute}, which does not exhibit bound state poles. While interfacial properties in the Blume-Capel model have been the subject of several Monte Carlo investigations (see in particular \cite{SY,SHK,AB,FS}), it is hard to distinguish numerically between interfacial wetting and weak binding, i.e. the formation of a bound state with mass (interfacial tension) $m_{+-}$ only slightly smaller than $2m_{+0}$. The sharp and general answer we are giving to this type of question provides a benchmark for future simulations. 

\begin{figure}[t]
\begin{center}
\includegraphics[width=12cm]{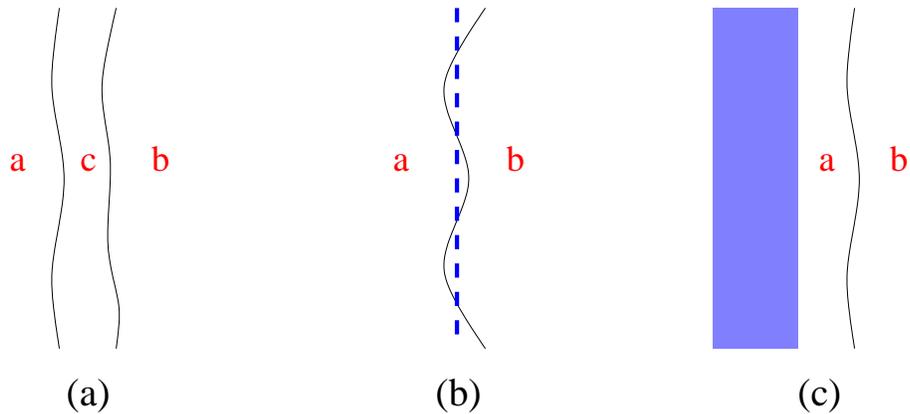}
\caption{Two interfaces enclose a third phase (a); an interface localized on a defect (b); phase enclosed between the boundary and an interface (c).}
\label{figure}
\end{center}
\end{figure}

\section{Localization on a defect line}
We now consider the bulk theory (\ref{bulk}) in presence of a defect line, and address the question of the localization of an interface by the defect. Using the notation $x=(x_1,x_2)$ for a point on the plane, the presence of the defect along the line $x_1=0$ corresponds to adding to (\ref{bulk}) the term
\EQ
{\cal H}_\textrm{defect}=-g\int d^2x\,\delta(x_1)\,\Psi(x)\,,
\label{defect}
\EN
where $\Psi(x)$ is an operator of the bulk theory with scaling dimension $X_\Psi$, so that $g$ has the dimension of a mass to the power $1-X_\Psi$. Within the one-dimensional quantum description with $x_2$ corresponding to imaginary time, the term (\ref{defect}) introduces an external potential centered at $x_1=0$ and vanishing as $x_1\to\pm\infty$. The problem of the interface in presence of the defect (Fig.~1b) maps to that of a kink in this potential. Here and in the rest of the paper the kinks remain excitations of the bulk theory (\ref{bulk}), so that (\ref{lambda}) and $m_{ab}\sim 1/\xi$ continue to hold\footnote{The defect also preserves the topological charge of the bulk ($g=0$) states; for neutral (charged) states the order parameter takes equal (different) values at $x_1=\pm\infty$.}.

If the defect is able to bind the kink, the energy of the bound state takes the form $E_\textrm{bound}=mf(z)<m$, where $f$ is a function of the dimensionless combination 
\EQ
z= g/m^{1-X_\Psi}\,,
\label{z}
\EN
and we drop indices on the kink mass $m$. The difference $1-f(z)$ measures the distance from the unbinding threshold, and then the strength of localization. The attractive regime is on one side, say $g>0$, of the non-interacting point $g=0$. As before, at the tail of renormalization group trajectories (large $m$) we can use the non-relativistic result that an attractive potential is binding, and conclude that the interface is localized for all positive $g$. But in turn this means localization for all positive $z$, and then also for small $m$. Binding vanishes as $g\to 0$, i.e. $f(z)\to 1$ as $z\to 0$. It then follows from (\ref{z}) that, as criticality is approached ($m\to 0$) for a fixed defect strength $g$, the localization strength vanishes if $X_\Psi>1$. 

For a defect realized as a line of weakened bonds, $\Psi$ coincides with the energy density operator $\varepsilon$, and we already quoted the values $X_\varepsilon=1$ and $4/5$ for the Ising and three-state Potts models, respectively. For the Ising model with this type of defect the absence of a delocalization transition was derived from the lattice in \cite{Abraham_defect}; it is also the only defect case, among those relevant for critical phenomena, which is exactly solvable \cite{DMS1,DMS2}. It has the peculiarity that $z=g$, so that the localization strength does not change as criticality is approached; the exact form of $f(g)$ can be deduced from the field theoretical solution of \cite{DMS1,DMS2}. 

 A different realization of the defect in the Ising model has been studied numerically in \cite{TAB}, where annealed vacancies were allowed along a line. Snapshots of the interface at temperatures sufficiently close to criticality show wide fluctuations for small vacancy density $\rho_D$, and clear localization on the defect for larger values of $\rho_D$.
 Within our framework, the Ising defect studied in \cite{TAB} corresponds to $g\sim\rho_D$ and $X_\Psi>2$. Indeed, since in the Ising model the only non-magnetic relevant operator is the energy density $\varepsilon$, the operator which creates the vacancies must be irrelevant; it is actually known \cite{Alyosha} that its scaling dimension is equal to 4. Since our analysis escludes a delocalization transition, the wide fluctuations of the interface at small $\rho_D$ must be interpreted as weak localization rather than delocalization. 

It was shown in \cite{AMW} for the lattice Ising model with a line of weakened bonds that depinning from the defect can be obtained inducing, through suitable boundary conditions, the interface to form (in average) an angle $\phi$ with the direction of the defect. Within our framework, this situation amounts to adding a left-right asymmetry (related to $\phi$) across the defect, which substantially modifies the analysis. Indeed, the non-relativistic limit now corresponds to an {\it asymmetric} potential well, which may or may not produce a bound state depending on the parameters of the well \cite{Landau}.

\section{Localization at a boundary}
We finally consider the problem of the localization of the interface by a boundary. We then consider the system on the half-plane $x_1\geq 0$, so that the reduced Hamiltonian is now given by (\ref{bulk}) (with the understanding that both ${\cal H}_\textrm{critical}$ and the integral are restricted to $x_1\geq 0$) plus a boundary term
\EQ
{\cal H}_\textrm{boundary}=\mu\int dx_2\,\Phi_B(x_2)\,,
\label{boundary}
\EN
where $\Phi_B(x_2)$ is a {\it boundary operator} with scaling dimension $X_{\Phi_B}$, and the coupling $\mu$ has dimension of a mass to the power $1-X_{\Phi_B}$. We also imply that the case $\mu=0$ corresponds to unconstrained, or {\it free}, boundary degrees of freedom. 

The truncation to $x_1\geq 0$ of the space on which the particles live no longer preserves the topological charge, so that interfaces can be created as the boundary parameter $\mu$ is varied. In particular, let us denote by $|0_a\rangle_B$ the states of the boundary system that for $\mu=0$ reduce to the degenerate ground states; the expectation value of the order parameter operator on these states tends for $x_1\to +\infty$ to the value $v_a$ it takes in the bulk phase $a$. Generically, the boundary field $\mu$ splits the degeneracy, and we indicate by $|0_a\rangle_B$ the ground state and by $|0_b\rangle_B$ one of the excited states. The latter corresponds to the ground state plus excitations, and a single kink excitation $a|b$ is allowed, since topological charge is not preserved. For $\mu$ small enough the order parameter in the excited state still tends to $v_b$ for $x_1\to\infty$, and the kink must be bound to the boundary to ensure this property (Fig.~1c). For the scaling Ising model with
  a boundary magnetic field such a bound state has been exhibited in \cite{GZ} as a pole of the exact kink reflection amplitude on the boundary; the kink unbinds when the field becomes strong enough. In general, the energy of the bound state is
\EQ
E_b=E_a+m_{ab}\cos\theta_0\,,
\label{young}
\EN
where $E_{a}$ (resp. $E_b$) is the energy of the state $|0_{a}\rangle_B$ (resp. $|0_b\rangle_B$), and the rapidity $\theta=i\theta_0$ of the bound kink is purely imaginary to make $E_b-E_a$ smaller than the unbinding threshold $m_{ab}$. As observed in \cite{contact1,contact2}, since $m_{ab}$ is the interfacial tension between phases $a$ and $b$, and $E_a$ (resp. $E_b$) the interfacial tension between the boundary and phase $a$ (resp. $b$), $\theta_0$ emerges as the {\it contact angle} of phenomenological wetting theory \cite{Dietrich,BEIMR}. $\theta_0$ is a function of the dimensionless combination
\EQ
s=\mu/m_{ab}^{1-X_{\Phi_B}}\,.
\label{s}
\EN
Binding is stronger when $E_b-E_a$ is small, i.e. for $\mu$ small, and consequently $s$ small. As $s$ increases binding weakens until a delocalization (or boundary wetting) transition takes place for a value $s_w$ at which $\theta_0(s_w)=0$. For $X_{\Phi_B}<1$, i.e. when the boundary operator is {\it relevant on the boundary}, it follows from (\ref{s}) that the interface will be delocalized sufficiently close to criticality ($m_{ab}\to 0$). 

The alternative scenario, i.e. $X_{\Phi_B}>1$ and binding growing as $m_{ab}$ decreases, namely when the interface fluctuates more, is not plausible. Understanding why it does not occur is instructive about the role of boundary operators in wetting phenomena. Consider for this purpose the case in which $\mu$ is a boundary magnetic field which for positive values favors phase $a$. It is then easy to see that the analysis we performed above applies only to the case of a relevant ($X_{\Phi_B}<1$) boundary operator. Indeed it requires that the boundary magnetization, which is responsible for the creation of the kink $a|b$ in $|0_b\rangle_B$ and goes as $\mu^{X_{\Phi_B}/(1-X_{\Phi_B})}$, is small for $\mu$ small, and then that $X_{\Phi_B}<1$. Notice that $\mu=0$ corresponds to free boundary spins, and $\mu=+\infty$ to boundary spins maximally polarized ({\it fixed}) in the direction $a$. These are scale (actually conformally \cite{Cardy}) invariant boundary conditions, and a boundary  magnetic operator relevant at $\mu=0$ induces a boundary renormalization group flow from free to fixed, i.e. towards the boundary fixed point with less degrees of freedom. This is what happens in the Ising model, where $X_{\Phi_B}$ equals 1/2 at the free boundary point and is larger than 1 at the fixed boundary point \cite{Cardy}.
On the other hand, a boundary magnetic operator irrelevant at $\mu=0$ requires the presence of an intermediate, partially polarized boundary fixed point at $\mu=\mu_*$, where it becomes relevant. The above analysis of the localization of the interface can then be repeated starting from this intermediate fixed point, replacing $\mu$ with $\mu-\mu_*$, and leads to the same conclusions. A partially polarized boundary fixed point is known to occur in the Blume-Capel model, for which $X_{\Phi_B}=3/2$ at $\mu=0$, and the vacancies make possible that the fixed point at $\mu_*$ has more degrees of freedom than that at $\mu=0$ \cite{Chim,Affleck}.

A situation different from that analyzed so far arises if $|0_a\rangle_B$ and $|0_b\rangle_B$ are states whose degeneracy is preserved by the boundary operator. In the Ising model, for example, modifying the boundary bond coupling from the bulk value $J$ to a value $J_0$ ($\mu\sim J-J_0$) preserves the degeneracy of $|0_+\rangle_B$ and $|0_-\rangle_B$. In such a case, for $m_{ab}$ large we can use the non-relativistic result that a potential well at the extremity of the half-line produces a bound state only if it is sufficiently deep \cite{Landau}. Hence the interaction, if attractive, localizes the interface for $s$ large enough, beyond a threshold $s_w$. Since, as already observed, for $\mu$ fixed localization cannot increase as $m_{ab}$ decreases, we conclude from (\ref{s}) that $X_{\Phi_B}>1$ at the free boundary condition point $\mu=0$. This prediction can be checked to be true in the Ising, Blume-Capel and Potts cases, for which boundary operators are classified \cite{Cardy,Chim,Affleck,AOS}. Putting all together, we see that localization at a boundary can follow different paths, but allows for the general conclusions that the interface delocalizes as bulk criticality is approached, and that the delocalized regime corresponds to an irrelevant boundary operator. 

The solution for the Ising model on the semi-infinite lattice was obtained in \cite{Abraham_wetting,Abraham} for fixed boundary spins {\it and} bonds coupling them to the adjacent spin column weakened from the bulk value $J$ to $J_0$. The overall effect is that of a boundary magnetic field varying from zero to infinity as $J_0$ varies from zero to $J$, and the solution exhibits the wetting transition at $s_w$ which, within our classification, falls into the energy splitting class.

In general, for fixed $\mu$, (\ref{lambda}) and (\ref{s}) select a wetting transition value $\lambda_w$ of the bulk parameter.  The relation 
\EQ
(1-\cos\theta_0)\propto(\lambda-\lambda_w)^{2-\alpha_S}
\label{alpha}
\EN
defines the interfacial (or surface) specific heat exponent $\alpha_S$ \cite{Dietrich,BEIMR}, and the transition is said to be continuous if $\alpha_S<1$. This denomination refers to the continuity of the first derivative of (\ref{alpha}) at $\lambda_w$, taking into account that the contact angle is phenomenologically set to zero in the delocalized regime $0<\lambda<\lambda_w$. It is important to realize, however, that analytically unbinding implies that the bound state pole in the kink-boundary scattering amplitude slides through a branch point into a second sheet of the complex energy plane \cite{Landau,ELOP}. Within the rapidity parameterization this means that the position $i\theta_0$ of the pole changes sign at $\lambda_w$, i.e.
\EQ
\theta_0\propto(\lambda-\lambda_w)^{2k+1}\,,\hspace{.7cm}k=0,1,2,\ldots\,,
\label{wetting}
\EN
so that 
\EQ
\alpha_S=-4k\,.
\EN
Eq.~(\ref{wetting}) also makes clear that $\alpha_S$ is not affected by renormalization, a feature that presumably persists in higher dimensions. A second exponent $\beta_S<0$ describes the divergence of the distance of the interface from the boundary (or wetting layer thickness) \cite{Dietrich,BEIMR}
\EQ
l\propto(\lambda-\lambda_w)^{\beta_S}\,.
\EN 
We have to identify $l$ with the inverse of the modulus $m_{ab}\theta_0$ ($\theta_0\ll 1$) of the momentum of the bound kink\footnote{The imaginary momentum causes the exponential decay of the wave function $e^{ipx_1}$ far from the boundary.}, so that 
\EQ
\beta_S=\alpha_S/2-1.
\EN 
Clearly, the value $k=0$ is the one expected in the generic case, and the associated values $\alpha_S=0$ and $\beta_S=-1$ indeed correspond to the Ising solution of \cite{Abraham,Abraham_wetting}. It will be interesting to establish in the future whether the values $k=1,2,\ldots$, which are also allowed by the theory, are realized in other universality classes and/or with different boundary conditions.

\section{Conclusion}
In this paper we showed how the theory of interface localization in near-critical planar systems with short range interactions can be formulated from first principles, without assuming models of the interface but obtaining instead its properties {\it within} the field theory associated to the given universality class of near-critical behavior. This allowed us to show, in particular, that the binding of two interfaces (and then interfacial wetting) is determined by the value of the bulk correlation length critical exponent. The strength of localization on a defect line may renormalize towards zero approaching criticality, but this is not sufficient to induce a delocalization transition. The latter occurs through different patterns in the boundary case, with the unifying feature that the boundary field becomes irrelevant in the delocalized regime. We showed that the surface specific heat and layer thickness exponents of the transition are not affected by renormalization and can take values that we determined exactly identifying the analytic mechanism underlying the wetting transition.

\end{document}